**IMPACT STUDIES USING A ONE STAGE LIGHT GAS GUN.** Jorge Carmona, Mike Cook, Jimmy Schmoke, Katie Harper, Jerry Reay, Lorin Matthews and Truell Hyde, CASPER (Center for Astrophysics, Space Physics and Engineering Research), Baylor University, Waco, TX 76798 E-mail: Truell_Hyde@baylor.edu.

**Introduction:** The Center for Astrophysics, Space Physics, and Engineering Research (CASPER) has completed construction and calibration of a Light Gas Gun (LGG), which is used for low velocity impact studies. At geosynchronous orbit, space debris can impact commercial satellites at velocities of 500 m/s [1] reducing their useful lifetime. Additionally, there is an ever-increasing population of abandoned non-operational satellites and related debris in these orbits [2]. Therefore, it is important to clearly understand the physics behind how such collisions can cause structural damage. This is most easily determined by measuring the damage incurred on representative material exposed to test collisions in the laboratory. Data collected in this manner will not only help illuminate the shock physics involved but can also aid in providing methods for designing advanced shielding for satellites.

**LGG:** The LGG employs a simple design incorporating interchangeable barrels of various bore diameters. The gun mount is adjustable for both azimuth and elevation, allowing an accuracy of ± 1.5 cm for impactor placement. Since the gun uses inert gas as a propellant, little maintenance is required and several shots per hour are possible. Barrels can easily be interchanged allowing the gun to fire spherical projectiles as small as 635 microns and as large as 2381.25 microns in diameter. The light gas gun operates by releasing high-pressure gas behind a projectile into a very low-pressure chamber (Figure 1). Currently, the gas used in the high-pressure chamber is either nitrogen or helium. Helium allows for higher velocities due to its lower molecular mass and is used the majority of the time. The low-pressure chamber is established using a vacuum system attached to the target end of the LGG. This system establishes a base pressure of 100 mTorr, removing most of the air molecules from inside the system and virtually eliminating air resistance for the projectile, thus increasing its final velocity. Projectile velocities of 783 ± 10 m/s are consistently recorded using 1000psi helium (2/3-pressure capacity of the system) while firing 1587.5 μm aluminum projectiles. Velocities as small as 115 m/s with an accuracy of ± 5 m/s can also be achieved using nitrogen. The theoretical upper bound for velocities attainable by the LGG are in the range of 1.0 km/s. There are several current limitations to the system. A much higher pressure must exist on the pressurized side of the valve than on the target chamber side of the valve for rough vacuum to be achieved in the system. This places a lower limit on the minimum velocity that can be achieved. A second limitation is the total number of shots that can be fired in a given configuration. Each piece in the LGG assembly is subject to wear, with the barrel capable of handling about one hundred shots before requiring replacement.

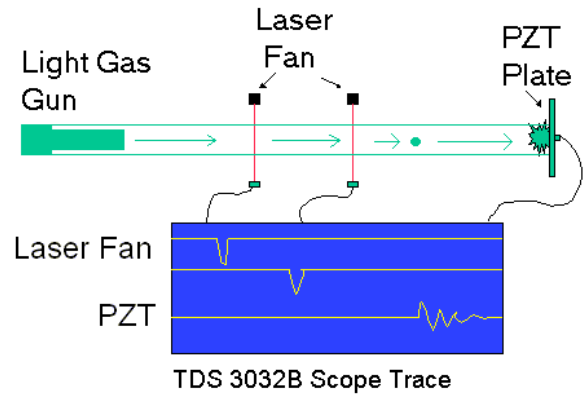

**Figure 1**. LGG Impact System. The barrel is contained within a pressurized chamber (in dark green); a valve allows the target chamber to be kept under vacuum (green arrows). As the particle travels through the chamber, it passes through two laser fans used to detect both velocity and particle size before striking the target plate to which is attached a PZT crystal which detects the impact.

**LGG Diagnostics:** System diagnostics consist of both optical and impact sensor devices. As shown in Fig. 2, projectile velocities are determined using a laser detection system capable of providing a presentable signal over a wide range of velocities and particle sizes. This laser diagnostic system is designed to register both the speed and pulse width of the change in the photon current, which in turn depend on the size and velocity of the particle. The measured signal must not exceed either the gain bandwidth product (GBP) or the slew rate (in volts/μs) of the amplifier. Since these are a measure of the amplifier's ability to change the output voltage with respect to input voltage, an amplifier with a large slew rate and high GBP is needed. The primary drawback to such high gain amplifiers is that the gain bandwidth is indirectly proportional to the maximum pulse width that can be amplified at a constant rate. (Gain bandwidth is defined as GBP/Acl where Acl is the amplifier's closed loop gain.) Once beyond the gain bandwidth, amplification is greatly reduced and decreases rapidly at a predetermined rate. To avoid this problem, several amplification stages are required so that the GBP of each stage remains within the desired pulse width regime. Since a single amplifier with extremely high gain would severely limit the overall bandwidth of the system, multiple cascade amplifiers are used. This design provides an overall system gain <40dBv with a bandwidth of 23 MHz and a gain flatness of 15 MHz with 0.1 dB deviation. The system has proven highly effective under actual laboratory conditions as a way to accurately analyze velocities of various sizes of particles. In the lower size limit, the capability of the electronics to record a significant amount of voltage gain is also necessary due to the almost negligible amount of photocurrent reduction created by the particle when passing through the beam. Since photon current is proportional to the electron current existing in the circuit, a relationship can be determined between the size of the particle and the amount of change in the photon current. The circuit employed is optimized for these conditions and analysis has shown that high gain amplification provides the amount of gain necessary to detect micron size particles. Preliminary circuit analysis has also shown that high gain amplification can be achieved without adversely affecting the fidelity of the output waveform.

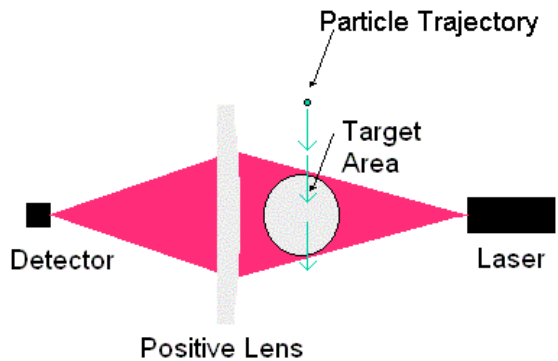

**Figure 2**. Side View of the Laser Fan. A top hat lens is used to fan the beam from a Lasiris SNF 670 nm Diode Laser, providing a large target area for the projectile (whose trajectory is into the page). The laser light is then focused by a second lens onto a Motorola Photo Detector MRD500 and the subsequent signal is measured by a Tektronix TDS 3032B oscilloscope.

The data from both laser fans as well as the PZT (piezoelectric lead zirconate titanate) crystal is recorded by two Tektronix TDS 3032B oscilloscopes, which are interfaced to a computer via a General Purpose Interface Bus (GPIB). The data is then analyzed using LabVIEW 6.1. A Fast Fourier Transform (FFT) analysis is conducted on the (PZT) data to determine the resonant frequency and the voltage output of the PZT plate. To minimize noise, a Butterworth filter is employed. The laser fan data is then passed through trigger levels that extract the information needed to calculate the velocity of the particle. Once the velocity and voltage response are known they are used to calculate the overall sensitivity of the plate. Finally, the raw data is saved for future analysis.

**Data and Analysis:** Currently the LGG is being used to conduct experiments on a stainless steel disk plate with a PZT crystal clamped at its center. Before the plate could be used with the LGG, it was first necessary to construct a sensitivity map (with sensitivity defined as peak-to-peak voltage divided by momentum in V/Ns) so that subsequent collected data could be properly understood. Such a map is extremely important since the sensitivity of the plate is greatly reduced for impacts displaced from the location of the PZT (in this case, the center of the plate). A properly calibrated sensitivity map can also help locate the specific impact point on the plate. After this map is obtained, the plate is again mapped using the LGG. The measured velocities from the PZT are then compared to known velocities from the optical diagnostics. The craters formed on the plate as well as the response of the PZT are both analyzed and then calibrated against particle velocity in order to provide the data necessary for gaining a better understanding of the potential damage to structures on orbit when subjected to space debris impacts. Specifically, data collected from the laser fan is collected using LabVIEW and then correlated to the PZT voltage response. It is then analyzed to find the relationship between the impactor diameter and velocity and the depth of the crater formed. The data shown in Figure 3 represents the results from the first 100 shots with the LGG system. As can be seen, nearly half (43) fell within the 5-7 V/Ns range. As expected, the data shows a linear relationship between particle velocity and voltage response from the PZT. The slope of the curve is used to determine the expected response for higher velocities. Particle velocity can also be determined given a known particle size and voltage response.

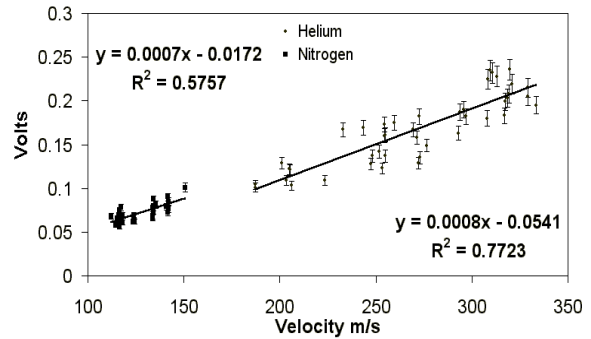

**Figure 3**. Shots with a sensitivity of 5-7 V/Ns

**Conclusion and Future Research:** A valid, low velocity light gas system has been developed and tested for use in orbital debris research. Future areas of research include employing additional PZTs to precisely locate the point of impact on the plate, higher velocity (hypervelocity) impacts of 1.0 to 2.0 km/s using a two stage light gas gun, and studies using aerogel as an advanced shielding material or as an interplanetary dust particle (IDP) capture mechanism.

**References:** [1] R. Westerkamp, J. Bendish, S. Theil and P. Wegener (1997) *Adv. Space Res.* **2**, p. 361-364. [2] M. Hechler (1985) *Adv. Space Res.* **5(2)**, p. 47-57.